\documentclass[aps,prl,floatfix,twocolumn]{revtex4-1}
\usepackage{epsfig}
\usepackage{verbatim}	 
\usepackage{amssymb}
\usepackage{bbold}
\usepackage{amsmath}
\usepackage{psfrag}
\usepackage{color}
\usepackage{ulem}
\newcommand{\bi}{\bibitem}
\newcommand{\Kf}{K\hspace{-0.8mm}f}
\usepackage[latin1]{inputenc}
\usepackage[english]{babel}
\usepackage{graphicx}
\usepackage{amscd}
\usepackage{eucal}
\usepackage{bm}
\usepackage{hyperref}
\input {epsf}
\begin{document}

\title{Robustness of Synchrony in Complex Networks and Generalized Kirchhoff Indices}
\author{M. Tyloo$^{1,2}$, T. Coletta$^{1}$, and Ph.~Jacquod$^{1}$}

\affiliation{$^1$School of Engineering, University of Applied Sciences of Western Switzerland HES-SO CH-1951 Sion, Switzerland \\
$^2$Institute of Physics, EPF Lausanne, CH-1015 Lausanne, Switzerland}

\date{\today}

\begin{abstract}
In network theory, a question of prime importance is how to assess network vulnerability in a fast and reliable manner. With this issue
in mind, we investigate the response to external perturbations of coupled dynamical systems on complex networks. We find that
for specific, non averaged perturbations, the response of synchronous states 
depends on the eigenvalues of the stability matrix of the unperturbed dynamics, as well as on its 
eigenmodes via their overlap with the perturbation vector. 
Once averaged over properly defined ensembles of perturbations, 
the response is given by new graph topological indices, which we introduce as {\it generalized 
Kirchhoff indices}. These findings allow for a fast and reliable method for assessing the specific or average
vulnerability of a network against changing operational conditions, faults or external attacks. 
\end{abstract}

\maketitle

{\bf Introduction.} Graph theory 
profoundly impacts many fields of human knowledge, including social and natural sciences, 
communication technology and electrical engineering, and information sciences and cybernetics ~\cite{Bar16}. Graphs allow for a 
convenient modelization of complex systems where their structure defines the couplings between the system's 
individual components, each of them with its 
own internal dynamics. The resulting coupled differential equations determine the system dynamics and its steady state 
solutions. Of particular interest is to predict the behavior of the system when it is perturbed away
from steady-state, for instance when an electric power plant goes offline in an operating power grid or when a line
is cut and information has to be redirected in a communication network. An issue of key importance for network security is how
to fast and reliably assess a network's vulnerability. This is not an easy task: network vulnerability depends on 
both the system dynamics and the network topology and 
geometry. It is highly desirable to identify a set of easily computed descriptors 
that characterize network vulnerability~\cite{Est10}. 
In this Letter we propose a new family of network descriptors in a two-step approach.
We investigate the sensitivity against external perturbations
of synchronous states of coupled dynamical systems on complex networks. First, we quantify this sensitivity using performance measures 
recently introduced in the context of electric power grids~\cite{Bam12,Bam14,Poo16}. Second,
by direct calculation of these performance measures, we identify a new class of easily computed topological indices that
generally characterize synchrony robustness or fragility under ensemble-averaged perturbations. 

Synchronization is ubiquitous~\cite{Stro04} in 
systems of coupled dynamical systems. It follows from the interplay between the internal dynamics of 
the individual systems and the coupling 
between them~\cite{Kur75,Ros98,Are08,Pik01}. Optimization of synchronization has been investigated from various angles.
The synchronous state can be optimal from the point of view of linear stability~\cite{Pec98}, the range of parameters that allow 
synchronization~\cite{Bar02,Cha05,Zho06}, the value that an order parameter takes at synchrony~\cite{Ska14} or the volume
of the basin of attraction around a stable synchronous fixed point~\cite{Wil06,Men13,Del17}. Here we extend 
these investigations by asking what makes synchronous states more or less fragile against external perturbations. 
For ensemble-averaged perturbations, the answer is surprisingly simple : 
synchrony fragility depends on a family of topological indices, which generalize
the Kirchhoff index introduced in Ref.~\cite{Kle93}. This result is rather general and remains valid for a large class of fragility performance 
measures quantifying the excursion away from the stable synchronous state, and for rather general synchronizing coupled dynamical systems.
Its main restriction is that it applies to 
not-too-large perturbations, which leave the system inside its original basin of stability. 

{\bf Model and method.} Our analysis focuses on the Kuramoto model~\cite{Kur75}
\begin{align}\label{eq:kuramoto}
 \dot{\theta}_i &= P_i - \sum_{j}b_{ij}\sin(\theta_i-\theta_j)\, , & i=1,...,n\, ,
\end{align}
though our results are more general and apply to a wider class of coupled dynamical systems (see Supplemental Material).
Eq.~(\ref{eq:kuramoto}) models the behavior of 
a set of $n$  harmonic oscillators, each with its angle coordinate $\theta_i$ and its natural frequency $P_i$,  
coupled to one another with couplings defined by the weighted adjacency matrix $b_{ij}\ge 0$. 
Kuramoto originally considered identical all-to-all coupling, $b_{ij}\equiv K/n$~\cite{Kur75}. 
It was found that for $K>K_c$, a finite number of oscillators synchronize, 
with $\dot{\theta}_i-\dot{\theta}_j=0$. This type of frequency synchronization also occurs for nonhomogeneous
couplings $b_{ij}$ defined on a complex network~\cite{Dor13}, the case of interest here.
Without loss of generality we set $\sum_i P_i=0$, for which
the frequency synchronous state has $\dot{\theta}_i \equiv 0$, $\forall i$~\cite{caveat1}. 

We consider a stable fixed-point solution ${\bm \theta}^{(0)}=(\theta_1^{(0)},\ldots ,\theta_n^{(0)}) $ to Eq.~(\ref{eq:kuramoto}) with
unperturbed natural frequencies $\bm{P}^{(0)}$. We then subject this state to a time-dependent perturbation
$\bm{P}(t) = \bm{P}^{(0)} + \delta \bm{P}(t)$, so that angles become time dependent, 
$\bm{\theta}(t) = \bm{\theta}^{(0)} + \delta \bm{\theta}(t)$. Linearizing the dynamics of Eq.~(\ref{eq:kuramoto}) about 
$\bm{\theta}^{(0)} $, one obtains  
\begin{align}\label{eq:kuramoto_lin}
 \delta \dot{\bm \theta} &= \delta {\bm P} - {\mathbb L}({\bm  \theta}^{(0)} ) \, \delta {\bm \theta} \, ,
\end{align}
where we introduced
the weighted Laplacian matrix ${\mathbb L}({\bm \theta^{(0)} })$ with matrix elements
\begin{equation}\label{eq:laplacian}
{\mathbb L}_{ij} = 
\left\{ 
\begin{array}{cc}
-b_{ij} \cos(\theta_i^{(0)} - \theta_j^{(0)}) \, , & i \ne j \, , \\
\sum_k b_{ik} \cos(\theta_i^{(0)} - \theta_k^{(0)}) \, , & i=j \, .
\end{array}
\right.
\end{equation}
This Laplacian is minus the stability matrix of the linearized dynamics, and 
since we consider a stable synchronous state, it
is positive semidefinite, with a single eigenvalue $\lambda_1=0$ with
eigenvector ${\bf u}_1=(1,1,1,...1)/\sqrt{n}$, and $\lambda_i>0$, $i=2,3,...n$.

The first term on the right-hand side of Eq.~(\ref{eq:kuramoto_lin}) perturbs angles away from the synchronous state. 
To assess the magnitude of this excursion in the spirit of Refs.~\cite{Bam12,Bam14,Poo16} we consider 
two fragility performance measures
\begin{subequations}\label{eq:c12}
\begin{eqnarray}\label{eq:c1}
{\mathcal C}_1(T) &=& \sum_i \int_0^T \, |\delta \theta_i(t) - \Delta(t) |^2 {\rm d}t  \, , \\ \label{eq:c2}
{\mathcal C}_2(T) &=& \sum_i  \int_0^T \, |\delta \dot{\theta}_i(t)- \dot \Delta(t) |^2 {\rm d}t   \, .
\end{eqnarray}
\end{subequations}
Because synchronous states are defined modulo any
homogeneous angle shift, the transformation $\theta_i^{(0)} \rightarrow \theta_i^{(0)} + C$ does not change the 
synchronous state. Accordingly, only angle shifts with $\sum_i \delta \theta_i(t)=0$ matter, which is incorporated
in the definitions of ${\mathcal C}_{1,2}$ by subtracting averages $\Delta (t) = n^{-1}
\sum_j \delta \theta_j(t)$ and $\dot\Delta (t)  = n^{-1} \sum_j \delta \dot\theta_j(t)$. 
An alternative procedure is to restrict oneself to perturbations orthogonal to ${\bf u}_1$~\cite{Bam12,Bam14,Poo16}.
Either procedure ensures, together with the non-negativity of ${\mathbb L}$, that ${\mathcal C}_{1,2}<\infty$, even when 
$T \rightarrow \infty$,
if the perturbation is short and weak enough that it leaves the dynamics inside the basin of attraction of 
$\bm{\theta}^{(0)}$. 
Low values for ${\mathcal C}_{1,2}^\infty \equiv {\mathcal C}_{1,2}(T\rightarrow \infty)$ indicate then that the system absorbs the perturbation with little fluctuations, while large values
indicate a temporary fragmentation of the system into independent pieces -- 
${\mathcal C}_{1,2}^\infty$ measures
the coherence of the synchronous state~\cite{Bam12}. 

We expand angle deviations over the eigenstates ${\bf u}_\alpha$ of $\mathbb L$,
$ \delta \bm{\theta}(t)=\sum_\alpha c_\alpha(t) \, {\bf u}_\alpha$, and rewrite Eq.~(\ref{eq:kuramoto_lin}) as 
\begin{equation}\label{eq:kuramoto_ca}
\dot{c}_\alpha(t) = \delta \bm{P}(t) \cdot {\bf u}_\alpha - \lambda_\alpha c_\alpha(t) \, ,
\end{equation}
whose general solution reads
\begin{equation}\label{eq:ca}
c_\alpha(t) = e^{-\lambda_\alpha t} c_\alpha(0) + e^{-\lambda_\alpha t} \, \int_0^t {\rm d} t' e^{\lambda_\alpha t'} \, \delta \bm{P}(t') \cdot {\bf u}_\alpha \, .
\end{equation}
Being interested in perturbations $\delta \bm{P}$ that start at $t=0$, when the system is in the synchronous state with $\delta \bm{\theta}(0)=0$, we set $c_\alpha(0) \equiv 0$. The performance measures of Eqs.~(\ref{eq:c12})
are given by ${\mathcal C}_1(T) = \sum_{\alpha\ge 2} \int_0^T \, c^2_\alpha(t) {\rm d}t$ and ${\mathcal C}_2(T) = \sum_{\alpha\ge 2} \int_0^T \, \dot{c}^2_\alpha(t) {\rm d}t$, as long as the perturbation is not too large so that the system eventually returns to its initial state. 
We next introduce generalized Kirchhoff indices
in terms of which we express ${\mathcal C}_{1,2}$ for three different classes of perturbations $\delta \bm{P}(t)$.

{\bf Generalized Kirchhoff indices.} 
The Kirchhoff index originally followed from the definition of the resistance distance in a graph~\cite{Kle93}.
To a connected graph, one associates an electrical 
network where each edge is a resistor given by the inverse edge weight in the original graph. 
The resistance distance is the resistance $\Omega_{ij}$ between any two nodes $i$ and $j$ on the electrical network. 
The Kirchhoff index is then defined as~\cite{Kle93} 
\begin{equation}\label{eq:kf1}
\Kf_1 \equiv \sum_{i<j} \Omega_{ij} \, , 
\end{equation}
where the sum runs over all pairs of nodes in the graph. 
For  a graph with Laplacian $\mathbb{L}$, it has been shown that $\Kf_1$ is given by 
the spectrum $\{ \lambda_\alpha\}$ of $\mathbb{L}$ as~\cite{Zhu96,Gut96,Col18}
\begin{equation}
\Kf_1 = n \, \sum_{\alpha \ge 2} \lambda_\alpha^{-1} \, .
\end{equation}
Up to a normalization prefactor,  $\Kf_1$ gives the mean resistance distance
$\overline{\Omega}$ over the whole graph. Intuitively, one expects the dynamics of a complex system 
to depend not only on $\overline{\Omega}$, but on the full set $\{ \Omega_{ij} \}$. Higher moments 
of $\{ \Omega_{ij} \}$ are encoded in generalized Kirchhoff indices $\Kf_m$ (see Supplemental Material) which we define as
\begin{equation}\label{eq:kirchhoff}
\Kf_m = n \, \sum_{\alpha \ge 2} \lambda_\alpha^{-m} \, ,
\end{equation}
for integers $m$. Below we show that ${\mathcal C}_{1,2}$ can be expressed as linear combinations of 
the $\Kf_m$'s corresponding to $\mathbb L$
in Eq.~(\ref{eq:laplacian}). 
We note that,
continued to $m \in {\mathbb C}$, 
$\Kf_m$ is known as the {\it spectral zeta function} of $\mathbb L$~\cite{Vor92}. 

{\bf Dirac delta perturbation.} We first consider $\delta \bm{P}(t) = \delta \bm{P}_0 \, \tau_0 \,
\delta(t)$ with the Dirac delta-function $\delta(t)$. Because the perturbation is limited in time, the limit $T \rightarrow \infty$ can be taken in 
Eqs.~(\ref{eq:c12}). One obtains (see Supplemental Material)  
\begin{subequations}\label{eq:c12deltan}
\begin{eqnarray}\label{eq:c1deltan}
{\mathcal C}_1^\infty &=&  \sum_\alpha \frac{(\delta \bm{P}_0 \cdot {\bf u}_\alpha)^2 \tau_0^2}{2} \,  \lambda_\alpha^{-1}\,  , \\ \label{eq:c2deltan}
{\mathcal C}_2^\infty &=&  \sum_\alpha \frac{(\delta \bm{P}_0 \cdot {\bf u}_\alpha)^2 \tau_0^2}{2} \, \lambda_\alpha  \, .
\end{eqnarray}
\end{subequations}
Both performance measures depend on the scalar product of the perturbation $\delta \bm{P}_0$ 
with the eigenmodes ${\bf u}_\alpha$ of $\mathbb L$. Such scalar products occur also when analyzing propagation of 
disturbances on networks~\cite{Ket16}.
To get more insight on 
 the typical network response, we define an ensemble of perturbation vectors with
 $\langle \delta P_{0i} \, \delta P_{0j}\rangle =\delta_{ij} \, \langle \delta P_{0}^2 \rangle$~\cite{caveat2}.
Averaging over that ensemble gives
\begin{subequations}\label{eq:c12delta}
\begin{eqnarray}\label{eq:c1delta}
\langle {\mathcal C}_1^ \infty \rangle &=&  \frac{\langle \delta P_{0}^2 \rangle \, \tau_0^2}{2n} \, \Kf_1 \,  , \\ \label{eq:c2delta}
\langle {\mathcal C}_2^\infty \rangle  &=&  \frac{\langle \delta P_{0}^2 \rangle \, \tau_0^2}{2n} \, \Kf_{-1}  \, .
\end{eqnarray}
\end{subequations}
The network structure determines the performance measures via the spectrum of the 
weighted Laplacian of Eq.~(\ref{eq:laplacian}).
The latter depends on the network structure -- its topology and edge weights, 
as well as the internal dynamics of the oscillators, which modifies the edge weights via angle differences 
$\theta_i^{(0)}-\theta_j^{(0)}$ determined by $\bm{P}^{(0)}$. The way all these ingredients determine average network fragility is
however simply encoded in $\Kf_{-1}$ and $\Kf_1$.
We note that  Eq.~(\ref{eq:c1delta}) appeared in slightly different, but equivalent form in 
Ref.~\cite{Bam12}.

{\bf Box perturbation.} Next, we 
go beyond the $\delta$ perturbations discussed so far~\cite{Bam12,Bam14,Poo16} and
consider a perturbation that is extended, but still limited in time, $\delta \bm{P}(t) = \delta \bm{P}_0 \,\Theta(t) \, \Theta(\tau_0-t)$, with the Heaviside function $\Theta(t) = 0$, $t<0$ and $\Theta(t) = 1$, $t>0$.  
Here also, the limit $T \rightarrow \infty$ can be taken in 
Eqs.~(\ref{eq:c12}). One obtains (see Supplemental Material)
\begin{subequations}\label{eq:c12heaviside}
\begin{eqnarray}\label{eq:c1heaviside} 
{\mathcal C}_1^\infty &=&   \sum_{\alpha\ge 2} \frac{(\delta \bm{P}_0 \cdot {\bf u}_\alpha)^2 }{\lambda_\alpha^3}
(\lambda_\alpha \tau_0-1+e^{-\lambda_\alpha \tau_0}) \, , 
\\ \label{eq:c2heaviside}
{\mathcal C}_2^\infty &=&  \sum_{\alpha\ge 2} \frac{(\delta \bm{P}_0 \cdot {\bf u}_\alpha)^2 }{\lambda_\alpha}
(1-e^{-\lambda_\alpha \tau_0}) \, .
\end{eqnarray}
\end{subequations} 
\begin{widetext}
As in Eqs.~(\ref{eq:c12deltan}), both performance measures depend on
$\delta \bm{P}_0 \cdot {\bf u}_\alpha$. 
 After averaging over the same ensemble of perturbation vectors as above,
 Eq.~(\ref{eq:c12heaviside}) becomes
(see Supplemental Material)
\begin{subequations}\label{eq:c12havg}
\begin{eqnarray}\label{eq:c1havg}
\langle {\mathcal C}_1^\infty \rangle&=&   \langle \delta P_0^2 \rangle \sum_{\alpha\ge 2} \frac{\lambda_\alpha \tau_0-1+e^{-\lambda_\alpha \tau_0}}{\lambda_\alpha^3}
\simeq \left \{
\begin{array}{c}
\langle \delta P_0^2 \rangle  \, \tau_0^2 \, \Kf_1 \big/ 2 n \, , \;\;\; \lambda_\alpha \tau_0 \ll 1 \, , \forall \alpha \,  , \\
\langle \delta P_0^2 \rangle \, \tau_0 \, \Kf_2 / n\, , \;\;\; \lambda_\alpha \tau_0 \gg 1  \, , \forall \alpha \, .
\end{array}
\right.  \\ \label{eq:c2havg}
\langle {\mathcal C}_2^\infty \rangle &=&  \langle \delta P_0^2 \rangle \sum_{\alpha\ge 2} \frac{1-e^{-\lambda_\alpha \tau_0}}{\lambda_\alpha}
\simeq \left \{
\begin{array}{c}
\langle \delta P_0^2 \rangle \, \tau_0 \, \Kf_0/n\, , \;\;\; \lambda_\alpha \tau_0 \ll 1 \, , \forall \alpha \,, \\
\langle \delta P_0^2 \rangle \, \Kf_1/n \, , \;\;\; \lambda_\alpha \tau_0 \gg 1  \, , \forall \alpha \,.
\end{array}  \right.
\end{eqnarray}
\end{subequations}
\end{widetext} 
Compared to Dirac delta perturbations,
$\langle \mathcal C_1^\infty \rangle$ now depends on $\Kf_2$ when $\tau_0$ is the longest time scale.
This is so, because time-extended perturbations
scatter through the network before they are damped by $\mathbb L$. Accordingly, they depend on details of the network
contained in higher moments of the distribution of resistance distances, hence on a generalized Kirchhoff index of higher order. 

{\bf Noisy perturbation.}
We finally consider fluctuating perturbations characterized by zero average and second moment
$\overline{ \delta P_i(t_1) \,
\delta P_j(t_2)} = \delta_{ij} \delta P_{0i}^2 \, \exp[- |t_1-t_2|/\tau_0]$ correlated over a typical time scale $\tau_0$.
Because this perturbation is not limited in time, we consider ${\mathcal C}_{1,2}(T)$ at finite but large $T$. 
Keeping only the leading-order term in $T$, we have (see Supplemental Material) 
\begin{subequations}\label{eq:c12noisy}
\begin{eqnarray}\label{eq:c1noisy}
\overline{ {\mathcal C}_1}(T)  &=&    T \, \sum_\alpha \frac{\sum_{i \in N_{\rm n}} \delta P^2_{0i} \, u^2_{\alpha,i}}{\lambda_\alpha (\lambda_\alpha+\tau_0^{-1})} + {\mathcal O}(T^0) \, , \\ \label{eq:c2noisy}
\overline{ {\mathcal C}_2}(T)  &=&      ( T/\tau_0) \, \sum_\alpha  \frac{\sum_{i \in N_{\rm n}} \delta P^2_{0i} \, u^2_{\alpha,i}}{\lambda_\alpha+\tau_0^{-1}} + {\mathcal O}(T^0) \, . \qquad
\end{eqnarray}
\end{subequations}
The response is determined by the overlap of the perturbation vector with the eigenmodes of $\mathbb L$.
The noise amplitude $\delta P_{0i}^2$ is localized on the set $N_{\rm n}$ of  noisy nodes. Averaging over an ensemble of perturbations
defined by all permutations of the noisy nodes over all nodes (see Supplemental Material),
$\langle {\mathcal C}_{1,2}\rangle$ is given by Eqs.~(\ref{eq:c12noisy}) with
$\sum_i \delta P_{0i}^2 u_{\alpha,i}^2 \rightarrow  \langle \delta P_0^2 \rangle$.  
If $\tau_0^{-1}$ lies inside the spectrum of $\mathbb L$, ${\cal C}_{1,2}$ are functions 
of the spectrum of $\mathbb L$ and the inverse correlation time $\tau_0^{-1}$.
If, on the other hand, $\tau_0^{-1}$ lies outside the spectrum of $\mathbb L$, averaged measures 
are directly expressable as infinite sums over generalized Kirchhoff indices, $\langle {\mathcal C}_{1,2}\rangle
= n^{-1} \, \langle \delta P_0^2 \rangle \, T \, \sum_{m=0}^{\infty} C_{1,2}^{(m)}$ with 
\begin{subequations}
\begin{eqnarray}
C_1^{(m)} &=&   
\left \{ 
\begin{array}{c}
(-1)^m \, \tau_0^{(m+1)} \Kf_{-m+1}  \, , \;\;\; \lambda_\alpha \tau_0 < 1 \, , \\ 
(-1)^m \, \tau_0^{-m} \Kf_{m+2}  \, , \;\;\; \lambda_\alpha \tau_0 > 1\, , 
\end{array}
\right. \\
C_2^{(m)} &=&   
\left \{ 
\begin{array}{c}
 (-1)^m \, \tau_0^{m} \Kf_{-m}  \, , \;\;\; \lambda_\alpha \tau_0 < 1\, , \\
 (-1)^m \, \tau_0^{-(m+1)} \Kf_{m+1}  \, , \;\;\; \lambda_\alpha \tau_0 > 1\, . 
\end{array}
\right.
\end{eqnarray}
\end{subequations}

{\bf Numerical Simulations.} To confirm our results numerically, we focus on ${\mathcal C}_1$ for both box and noisy perturbations,
varying their time scale $\tau_0$. 
We consider Eq.~(\ref{eq:kuramoto}) with two types of networks, (i) small-world networks, 
where a cycle graph with constant coupling $b_{ij}=b_0$ for any node $i$ to 
its $4$ nearest neighbors undergoes random rewiring with probability $p \in [0,1]$~\cite{Wat98,caveat3}, and
(ii) simple cyclic networks where each node is coupled to its nearest- and $q^{\rm th}$-neighbors with a constant coupling $b_{i,i\pm1}=b_{i,i\pm q}=b_0$ (see inset in Fig.~\ref{fig:fig2}). 
In both cases, we fix the number of nodes to $n=50$. In all cases, the unperturbed natural frequencies vanish, $P_i^{(0)}=0$.
The box perturbation 
has $\delta \bm{P}_0= (0,0,..., \delta P_{0i_1},0,...,\delta P_{0i_2},0,...)$ with $\delta P_{0i_1}=-\delta P_{0i_2}=0.01 \, b_0$,
and averaging is performed over all pairs of nodes $(i_1,i_2)$. The noisy perturbation acts on all nodes, and we construct 
noise sequences $P_i(t)$ satisfying $\overline{  \delta P_i(t_1) \,
\delta P_j(t_2) }= \delta_{ij} \delta P_{0i}^2 \, \exp[- |t_1-t_2|/\tau_0]$ using the method described in Ref.~\cite{Fox88},
with $\delta P_{0i}=0.01 \, b_0$.

\begin{figure}
\centering
\hspace{-3mm}\includegraphics[width=1.02 \columnwidth]{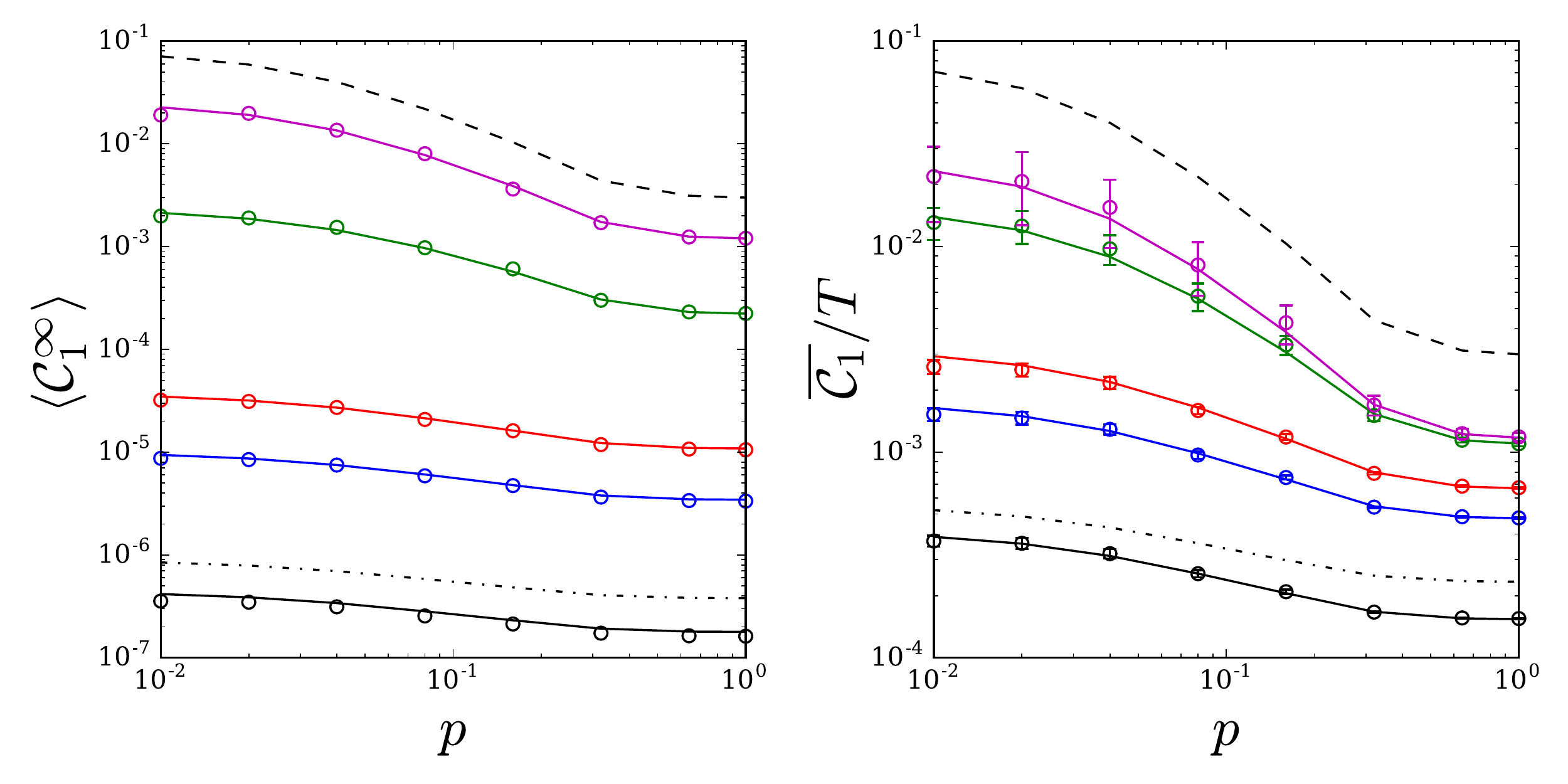}
\caption{\label{fig:fig1} (Color online) Performance measure $\langle {\mathcal C}_1^\infty \rangle$ (for box perturbation, left panel)
and $\overline{ {\mathcal C}_1}/T$ (for noisy perturbation, right panel)
for the small-world model with $n=50$ nodes as a function of the rewiring probability $p$~\cite{Wat98} and with
$\tau_0=0.1/b_0$ (black), $0.5/b_0$ (blue), $1/b_0$ (red), $10/b_0$ (green) and $50/b_0$ (violet). Solid lines give Eqs.~(\ref{eq:c1havg})
(left) and (\ref{eq:c1noisy}) (right) calculated numerically over an ensemble of networks obtained from 20 different rewirings. 
The dotted-dashed lines give $\Kf_1$ and the dashed lines $\Kf_2$, both vertically shifted.
In the right panel, $\overline{ {\mathcal C}_1}(T)$ is averaged over $T' \in [T-200/b_0,T+200/b_0]$ with $T=800/b_0$, and
error bars give the standard deviation of numerically obtained values with 10 different noise sequences.\\[-9mm]}
\end{figure}

The theory is numerically 
confirmed for small-world networks in Fig.~\ref{fig:fig1}, where ${\cal C}_1$ decreases monotonically as the 
rewiring probability $p$ increases, in complete agreement with 
Eqs.~(\ref{eq:c1havg}) and (\ref{eq:c1noisy}) (colored solid lines). This is qualitatively understood as follows.
As $p$ increases and more network edges are rewired, more couplings with longer range appear in the network, which stiffens the 
synchronous state. Fig.~\ref{fig:fig1} shows that the resulting decrease in fragility of synchrony occurs already with $p \simeq 0.1-0.2$,
where only few long-range couplings exist in the network -- true small-world networks~\cite{Wat98}.
Earlier works showed that small-world networks have larger range of parameters over which synchrony prevails, compared to
random networks~\cite{Bar02}. Fig.~\ref{fig:fig1} shows that, additionally, synchronous states in
small-world networks are more robust than in regular networks.

Further insight into synchrony fragility 
is obtained when considering our cyclic graph model with nearest- and $q^{\rm th}$-neighbor coupling. 
If the range of the coupling were the only ingredient determining the fragility of the synchronous state, then 
one would observe a monotonic decrease of ${\cal C}_1$ as a function of $q$. 
Fig.~\ref{fig:fig2} shows numerical results for the cyclic graphs and five values of $\tau_0$ ranging from 
$\lambda_\alpha \tau_0 \lesssim 1$ to $\lambda_\alpha \tau_0 \gtrsim 1$, $\forall \alpha$. Analytical results of Eqs.~(\ref{eq:c1havg}) 
and (\ref{eq:c1noisy}), in particular, the crossover from $\langle {\cal C}_1^\infty \rangle \sim \Kf_1$ to  $\langle {\cal C}_1^\infty \rangle \sim \Kf_2$ predicted in Eq.~(\ref{eq:c1havg}) 
when $\tau_0$ increases,
are clearly confirmed. Particularly remarkable is that $\Kf_1$ and $\Kf_2$ are not monotonous in 
the coupling range $q$ (see Supplemental Material), 
which is clearly reflected in the behavior of $\langle {\cal C}_1^\infty \rangle$. This unambiguously demonstrates that 
average fragility of
synchrony does not depend trivially on the range of the couplings between oscillators, but is entirely determined 
by generalized Kirchhoff indices.

\begin{figure}
\centering
\hspace{-3mm}\includegraphics[width=1.02 \columnwidth]{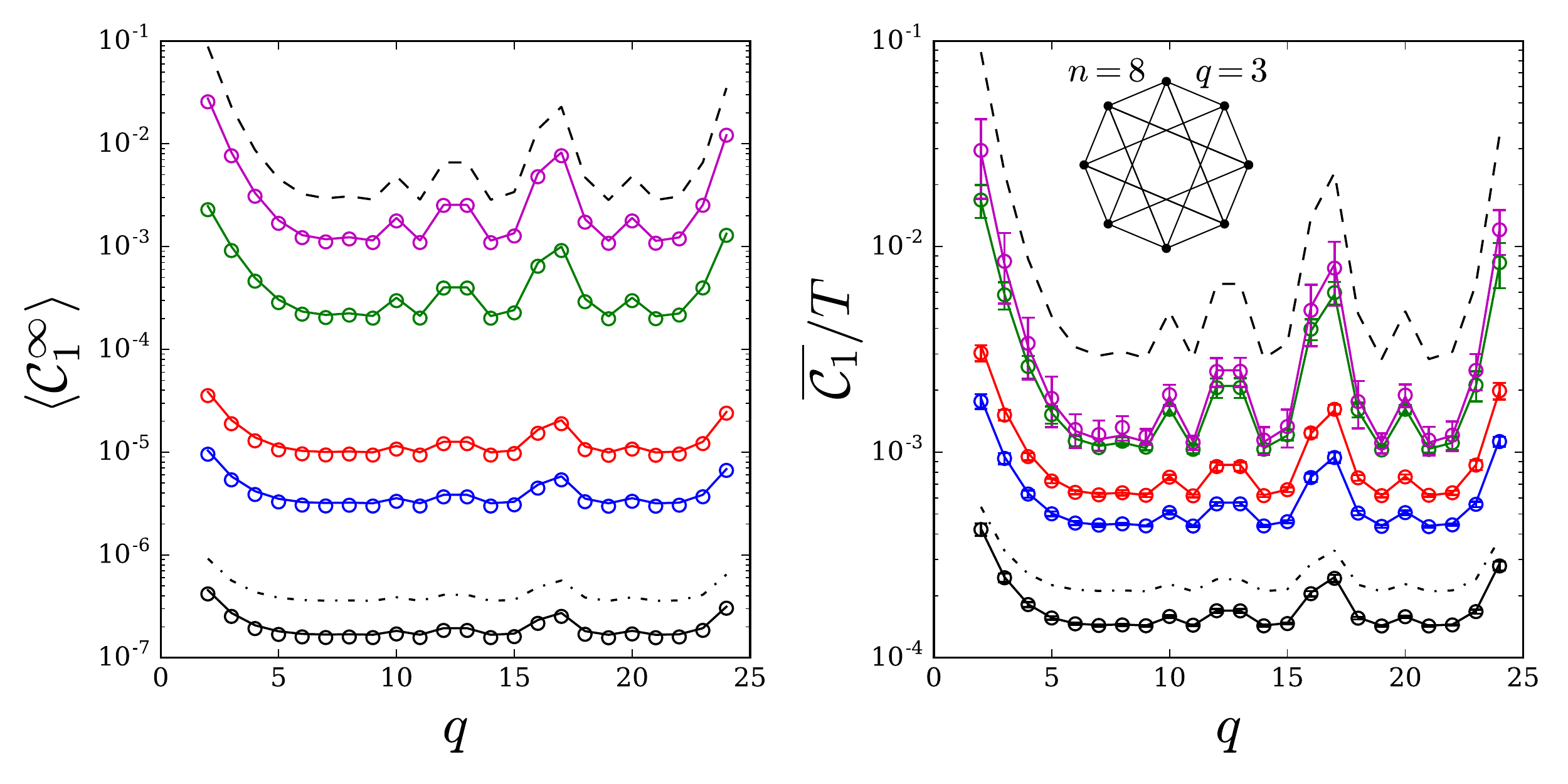}
\caption{\label{fig:fig2} (Color online) Performance measure $\langle {\mathcal C}_1^\infty \rangle$ (for box perturbation, left panel)
and $\overline{ {\mathcal C}_1}/T$ (for noisy perturbation, right panel)
for the cyclic graph with $n=50$ nodes with nearest- and $q^{\rm th}$-neighbor coupling, $b_{i,i\pm1}=b_{i,i\pm q}=b_0$,
as a function of $q$ and with
$\tau_0=0.1/b_0$ (black), $0.5/b_0$ (blue), $1/b_0$ (red), $10/b_0$ (green) and $50/b_0$ (violet). Solid lines give Eqs.~(\ref{eq:c1havg})
(left) and (\ref{eq:c1noisy}) (right). The dotted-dashed lines give $\Kf_1$ and the dashed lines $\Kf_2$, both vertically shifted.
In the right panel, $\overline{ {\mathcal C}_1}(T)$ is averaged over $T' \in [T-200/b_0,T+200/b_0]$ with $T=800/b_0$, and
error bars give the standard deviation of numerically obtained values with 10 different realizations of
noisy perturbations.
The inset sketches the model for $n=8$ and $q=3$.\\[-9mm]}
\end{figure}

{\bf Conclusion.} Using both performance measures defined in Eqs.~(\ref{eq:c12}),
we have expressed synchrony fragility in terms of
the weighted Laplacian matrix $\mathbb L$ of the system's network. We have first shown that
the response to specific perturbations is determined by both the spectrum of $\mathbb L$ and
its eigenmodes ${\bf u}_\alpha $ through their scalar product $\delta \bm{P}_0 \cdot {\bf u}_\alpha $ with the perturbation vector.
Eqs.~(\ref{eq:c12deltan}), (\ref{eq:c12heaviside}) and (\ref{eq:c12noisy}) clearly indicate that perturbations overlapping with 
the eigenmodes with smallest Lyapunov exponents have the largest impact on the synchronous state. The most vulnerable nodes
are accordingly identified as the nodes carrying these eigenmodes. Second, we considered
performance measures averaged over ergodic ensembles of perturbations. In this case, they depend on $\mathbb L$ only through 
generalized Kirchhoff indices, which we introduced in Eq.~(\ref{eq:kirchhoff}). The latter are both spectral and 
topological in nature, as they can be reexpressed in terms of the resistance distances in the virtual network defined by $\mathbb L$ (see Supplemental Material). A network's average or global fragility can therefore be easily quantified 
by a direct calculation of generalized Kirchhoff indices. This is a 
computationally easy task, requiring in most instances to determine few of the smallest eigenvalues of $\mathbb L$, and that, for a given
system, can be done for few typical fixed points once and for all.
Our findings are rather general and generalized Kirchhoff indices naturally characterize the fragility of synchronous states 
for many coupled dynamical systems, both beyond the Kuramoto model considered here as well as for other types of perturbation not
discussed here (see Supplemental Material). 

Two extensions of this work should be considered. First,
our approach has been based on the implicit assumption that the perturbation is sufficiently weak, such that the system 
stays close to its initial state. Criteria for acute vulnerability should account for the breakdown of this assumption and
quantify the perturbation threshold above which networks either lose synchrony or 
change their synchronous state.
Second, synchrony fragility for second-order systems with inertia should be considered, investigating in particular more 
closely the case of electric power grids under the influence of fluctuating power injections~\cite{Zha17}. Work along those lines is in progress.

This work has been supported by the Swiss National Science Foundation under an AP Energy Grant.

\pagebreak
\onecolumngrid
\begin{center}
\textbf{\large Robustness of Synchrony in Complex Networks and Generalized Kirchhoff Indices : Supplemental Material}
\end{center}
\setcounter{equation}{0}
\setcounter{figure}{0}
\setcounter{table}{0}
\makeatletter
\renewcommand{\theequation}{S\arabic{equation}}
\renewcommand{\thefigure}{S\arabic{figure}}
\renewcommand{\bibnumfmt}[1]{[S#1]}
\renewcommand{\citenumfont}[1]{S#1}

\section{Generalized Kirchhoff indices}
For a complex graph determined by its Laplacian matrix $\mathbb L$, we introduce a family of graph invariants
\begin{eqnarray}
\Kf_m=n\sum_{\alpha\ge 2}\lambda_{\alpha}^{-m} \, ,
\end{eqnarray}
where $\lambda_\alpha$ is an eigenvalue of $\mathbb L$. We call them {\it generalized Kirchhoff indices} because
the Kirchhoff index introduced in Ref.~\cite{Kle93} can be expressed as~\cite{Zhu96,Gut96}
\begin{eqnarray}
\Kf_1=n\sum_{\alpha\ge 2}\lambda_{\alpha}^{-1}\, .
\end{eqnarray}
We show that, just like the original Kirchhoff index $\Kf_1$, generalized Kirchhoff 
indices can be expressed as functions of the resistance distances between any pair of nodes $(i,j)$ in the network.
The network's Laplacian matrix $\mathbb L$ has one zero eigenvalue. We therefore define 
the matrix ${\bm \Gamma}$ 
\begin{equation}
 {\bm \Gamma}={\mathbb L} + {{\bf u}_{1}}^\top {\bf u}_{1}\,,
\end{equation}
in terms of which the resistance distance $\Omega_{ij}$ between nodes $i$ and $j$ is defined as~\cite{Kle93}
\begin{equation}\label{eq:Resistance distance}
 \Omega_{ij}=\Gamma_{ii}^{-1}+\Gamma_{jj}^{-1}-2\Gamma_{ij}^{-1}\,  .
\end{equation}
This can be rewritten in terms of the eigenvectors of $\mathbb L$ as~\cite{Col17}
\begin{eqnarray}
\Omega_{ij}=\sum_{\alpha\ge 2}\frac{(u_{\alpha, i}-u_{\alpha, j})^2}{\lambda_\alpha} \, ,
\end{eqnarray}
where the zero mode corresponding to $\lambda_1=0$ is omitted in the sum.
We show that $\Kf_1$ and $\Kf_2$ can be rewritten in terms of resistance distances. For $\Kf_1$, one has
\begin{eqnarray}
\sum_{i<j}\Omega_{ij}&=&\frac{1}{2}\sum_{i,j} \, \sum_{\alpha\ge 2}\frac{(u_{\alpha, i}-u_{\alpha, j})^2}{\lambda_\alpha} =  \Kf_1 \, ,
\end{eqnarray}
because the eigenvectors $\alpha\ge2$ of $\mathbb L$ satisfy $\sum_i u_{\alpha, i}=0$ and $\sum_i u_{\alpha, i}^2=1$. 
To express $\Kf_2$, higher moments of the distribution of resistance distances are needed. One has, 
\begin{eqnarray}
\sum_{i,j}\Omega_{ij}^2&=&\sum_{i,j\, ;\, \alpha,\beta \ge 2}\frac{(u_{\alpha, i}-u_{\alpha, j})^2(u_{\beta, i}-u_{\beta, j})^2}{\lambda_\alpha\lambda_{\beta}} = 2n\sum_{i\,;\,\alpha,\beta\ge 2}\frac{u_{\alpha, i}^2 u_{\beta, i}^2}{\lambda_\alpha \lambda_\beta} + \frac{2(\Kf_1)^2}{n^2} + \frac{4 \Kf_2}{n}\, ,  \qquad
\\
\sum_{i,j,k}\Omega_{ij}\Omega_{jk}&=&\sum_{i,j,k\,;\,\alpha, \beta \ge 2}\frac{(u_{\alpha, i}-u_{\alpha, j})^2(u_{\beta, j}-u_{\beta, k})^2}{\lambda_\alpha\lambda_{\beta}} = \frac{3(\Kf_1)^2}{n} + n^2\sum_{i\,;\,\alpha,\beta\ge 2}\frac{u_{\alpha, i}^2 u_{\beta, i}^2}{\lambda_\alpha \lambda_\beta} \, .
\end{eqnarray}
Combining the latter two equations, one has
\begin{eqnarray}
\frac{n}{4}\left(\sum_{i,j}\Omega_{ij}^2\right)-\frac{1}{2}\left(\sum_{i,j,k}\Omega_{ij}\Omega_{jk}\right) + \frac{(\Kf_1)^2}{n} = \Kf_2 \, .
\end{eqnarray}
For $m\ge 3$, it is possible though algebraically tedious to show that 
$\Kf_m$  can be expressed in a similar way in terms of higher moments of resistance distances.

\section{Direct calculation of fragility measures}
The fragility performance measures introduced in Eqs.~(4a,b) of the main text can be rewritten in terms of the coefficients
of the expansion $\delta\vec{\theta}(t)=\sum_\alpha c_{\alpha}(t){\bf u}_\alpha$
of angle displacements over the eigenvectors ${\bf{u}_\alpha}$ of $\mathbb L$. One has 
\begin{eqnarray}
{\mathcal C}_1(T) = \sum_{\alpha\ge2}\int_0^Tc_\alpha^2(t){\rm d} t \; , \;\;\;\,\;\;\;\,\;
{\mathcal C}_2(T) =\sum_{\alpha\ge2}\int_0^T\dot{c}_\alpha^2(t){\rm d} t\label{eq:c12} \, .
\end{eqnarray}
The coefficients $c_\alpha(t)$ are solutions of
\begin{equation}\label{eq:ca}
c_\alpha(t) = e^{-\lambda_\alpha t} c_\alpha(0) + e^{-\lambda_\alpha t} \, \int_0^t {\rm d} t' e^{\lambda_\alpha t'} \, \delta {\bm{P}}(t') \cdot {\bf u}_\alpha \, ,
\end{equation}
and in our case where the perturbation starts at $t=0$, $c_\alpha(0)=0$. We treat sequentially, and
in some additional details, the three perturbations 
considered in the main text. 

\subsection{Dirac delta perturbation}
We first consider $\delta {\bm P}(t)=\delta{\bm P}_0\tau_0\delta(t)$. Inserting it into Eq.~(\ref{eq:ca}) one obtains,
\begin{eqnarray}
c_{\alpha}(t)=(\delta{\bm P}_0\cdot {\bf u}_\alpha) \tau_0 e^{-\lambda_{\alpha}t}.
\end{eqnarray}
This directly gives
\begin{eqnarray}
{\mathcal C}_1(T)=\sum_{\alpha\ge2}\frac{(\delta {\bm P}_0 \cdot {\bf u}_\alpha)^2\tau_0^2}{2\lambda_\alpha}(1-e^{-2\lambda_\alpha T})
 \; , \;\;\;\,\;\;
{\mathcal C}_2(T)=\sum_{\alpha\ge2}\frac{(\delta {\bm P}_0 \cdot {\bf u}_\alpha)^2\tau_0^2}{2}\lambda_\alpha(1-e^{-2\lambda_\alpha T})\, . \qquad
\end{eqnarray}
Taking the limit $\lambda_\alpha T \gg 1$, one obtains Eqs.~(10a,b) of the main text,
\begin{subequations}
\begin{eqnarray}
{\mathcal C}_1(T\rightarrow\infty)&=&{\mathcal C}_1^\infty=\frac{\tau_0^2}{2}\sum_{\alpha\ge2} \frac{(\delta {\bm P}_0 \cdot {\bf u}_\alpha)^2}{\lambda_\alpha} \, , \\
{\mathcal C}_2(T\rightarrow\infty)&=&{\mathcal C}_2^\infty=\frac{\tau_0^2}{2}\sum_{\alpha\ge2} (\delta {\bm P}_0 \cdot {\bf u}_\alpha)^2\lambda_\alpha \, .
\end{eqnarray}
\end{subequations}
Averaging over the ensemble of perturbation vectors defined by $\langle \delta P_{0i} \rangle=0$, $\langle \delta P_{0i}\delta P_{0j} \rangle=\delta_{ij}\langle \delta P_0^2 \rangle$, we have $\langle (\delta {\bm P}_0 \cdot {\bf u}_\alpha)^2\rangle = 
\sum_{i,j}\langle  \delta P_{0i} \delta P_{0j} \rangle \,  {\bf u}_{\alpha,i}  {\bf u}_{\alpha,j} =
\langle \delta P_0^2 \rangle $, $\forall \alpha$. The averaged fragility measures become,
\begin{subequations}
\begin{eqnarray}
\langle {\mathcal C}_1^\infty \rangle &=&\frac{\langle\delta P_0^2 \rangle\tau_0^2}{2}\sum_{\alpha\ge2} \frac{1}{\lambda_\alpha}=\frac{\langle\delta P_0^2 \rangle\tau_0^2}{2n}\Kf_1 \, , \\
\langle {\mathcal C}_2^\infty \rangle &=&\frac{\langle\delta P_0^2 \rangle\tau_0^2}{2}\sum_{\alpha\ge2} \lambda_\alpha=\frac{\langle\delta P_0^2 \rangle\tau_0^2}{2n}\Kf_{-1} \, ,
\end{eqnarray}
\end{subequations}
just as in Eqs.~(11a,b) in the main text.

\subsection{Box perturbation}
We next consider $\delta {\bm P}(t)=\delta{\bm P}_0\Theta(t)\Theta(\tau_0-t)$. Eq.~(\ref{eq:ca}) with $c_\alpha(0)=0$ now gives,
\begin{eqnarray}
c_{\alpha}(t)=\left \{
\begin{array}{c}
(\delta {\bm P}_0 \cdot {\bf u}_\alpha)(1-e^{-\lambda_\alpha t})\big/\lambda_\alpha \, , \;\;\;  t \le \tau_0 \, , \\
(\delta {\bm P}_0 \cdot {\bf u}_\alpha)(e^{\lambda_\alpha(\tau_0-t)}-e^{-\lambda_\alpha t})\big/\lambda_\alpha \, , \;\;\;  t > \tau_0 \, . 
\end{array}
\right.
\end{eqnarray}
Eqs.~(\ref{eq:c12}) become
\begin{subequations}
\begin{eqnarray}
{\mathcal C}_1(T)&=&\sum_{\alpha\ge2}\frac{(\delta {\bm P}_0 \cdot {\bf u}_\alpha)^2}{\lambda_\alpha^3}(\lambda_\alpha \tau_0 -1 +e^{-\lambda_\alpha \tau_0} - \frac{e^{2\lambda_\alpha(\tau_0-T)}}{2}+e^{\lambda_\alpha(\tau_0-T)}-\frac{e^{-2\lambda_\alpha T}}{2}) \, , \\
{\mathcal C}_2(T)&=&\sum_{\alpha\ge2}\frac{(\delta {\bm P}_0 \cdot {\bf u}_\alpha)^2}{\lambda_\alpha}(1-e^{-\lambda_\alpha \tau_0}-\frac{e^{2\lambda_\alpha(\tau_0-T)}}{2}+e^{\lambda_\alpha(\tau_0-2T)}-\frac{e^{-2\lambda_\alpha T}}{2})\, .
\end{eqnarray}
\end{subequations}
Taking the limit $\lambda_\alpha T \gg 1$, one recovers Eqs.~(12a,b) in the main text,
\begin{subequations}
\begin{eqnarray}
{\mathcal C}_1^\infty&=&\sum_{\alpha\ge2}\frac{(\delta {\bm P}_0 \cdot {\bf u}_\alpha)^2}{\lambda_\alpha^3}(\lambda_\alpha \tau_0 -1 +e^{-\lambda_\alpha \tau_0}) \, , \\
{\mathcal C}_2^\infty&=&\sum_{\alpha\ge2}\frac{(\delta {\bm P}_0 \cdot {\bf u}_\alpha)^2}{\lambda_\alpha}(1-e^{-\lambda_\alpha \tau_0})\, .
\end{eqnarray}
\end{subequations}
Following the same averaging procedure as for the $\delta$-perturbation, one finally recovers Eqs.~(13a,b) in the main text,
\begin{subequations}
\begin{eqnarray}
\langle {\mathcal C}_1^\infty \rangle &=&\langle \delta P_0^2 \rangle \sum_{\alpha\ge2}\frac{1}{\lambda_\alpha^3}(\lambda_\alpha \tau_0 -1 +e^{-\lambda_\alpha \tau_0}) \, , \\
\langle{\mathcal C}_2^\infty \rangle &=&\langle \delta P_0^2 \rangle\sum_{\alpha\ge2}\frac{1}{\lambda_\alpha}(1-e^{-\lambda_\alpha \tau_0})\, .
\end{eqnarray}
\end{subequations}
The asymptotic behaviors for $\lambda_\alpha\tau_0\ll 1$ and $\lambda_\alpha\tau_0\gg 1$ are easily computed via a 
Taylor-expansion.

\subsection{Noisy perturbation}
We finally consider fluctuating perturbations characterized by zero average $\overline{\delta P_{0i}}=0$, and second moments $\overline{\delta P_i(t_1) \, \delta P_j(t_2)} = \delta_{ij} \delta P_{0i}^2 \, \exp[-\tau_0^{-1} |t_1-t_2|]$ correlated over a typical time scale $\tau_0$. 
With this ensemble average, one obtains,
\begin{eqnarray*}
\overline{{\mathcal{C}_1}}(T) &=& \sum_{\alpha\ge 2}\int_0^T \ \overline{ c_{\alpha}^2(t)} {\rm d} t\\
&=& \sum_{\alpha\ge 2}\int_0^T e^{-2\lambda_\alpha t}\int_0^t \int_0^t e^{\lambda_\alpha(t_1+t_2)} \, \overline{ \delta {\bm P}(t_1) \cdot {\bf u}_\alpha\, \delta {\bm P}(t_2) \cdot {\bf u}_\alpha } \,  {\rm d} t_1{\rm d} t_2{\rm d} t\\
&=&\sum_{\alpha\ge 2}\sum_i (\delta P_{0i} {u}_{\alpha, i})^2\int_0^Te^{-2\lambda_\alpha t}\int_0^t \int_0^t e^{\lambda_\alpha(t_1+t_2)}e^{-|t_1-t_2|/\tau_0} \, {\rm d} t_1 \, {\rm d} t_2 \, {\rm d} t\\
&=&\sum_{\alpha\ge 2}\sum_i (\delta P_{0i} {u}_{\alpha, i})^2\left[ \frac{T}{\lambda_\alpha(\lambda_\alpha + \tau_0^{-1})} + \frac{1-e^{-2\lambda_\alpha T}}{2\lambda_\alpha^2(\lambda_\alpha - \tau_0^{-1})} + \frac{2(e^{-(\lambda_\alpha + \tau_0^{-1})T}-1)}{(\lambda_\alpha+\tau_0^{-1})(\lambda_\alpha^2 - \tau_0^{-2})} \right].
\end{eqnarray*}
To calculate $\overline{{\cal C}_2}$ we note that,
\begin{eqnarray}
\overline{\dot{c}_{\alpha}^2}(t) &=&\lambda_\alpha^2 \overline{{c}_{\alpha}^2}(t)  + \overline{(\delta {\bm P}(t) \cdot {\bf u_{\alpha}})^2} - 2\lambda_\alpha e^{-\lambda_\alpha t}\int_0^te^{\lambda_\alpha t'} \, \overline{\delta {\bm P}(t') \cdot {\bf u}_\alpha\, \delta {\bm P}(t) \cdot {\bf u}_\alpha} \,  {\rm d} t' \, .
\end{eqnarray}
One obtains, after some algebra
\begin{eqnarray*}
\overline{{\mathcal{C}_2}}(T) \rangle &=&\sum_{\alpha\ge 2}\sum_i (\delta P_{0i} {u}_{\alpha, i})^2\left[ \frac{T}{\tau_0 (\lambda_\alpha + \tau_0^{-1})} + \frac{1-e^{-2\lambda_\alpha T}}{2(\lambda_\alpha - \tau_0^{-1})} + \frac{2\lambda_\alpha\tau_0^{-1}(e^{-(\lambda_\alpha + \tau_0^{-1})T}-1)}{(\lambda_\alpha+\tau_0^{-1})(\lambda_\alpha^2 - \tau_0^{-2})} \right].
\end{eqnarray*}
For $\lambda_\alpha T \gg 1$, one recovers Eqs.~(14a,b) in the main text,
\begin{eqnarray}
\overline{{\mathcal{C}_1}}(T)  &=&\sum_{\alpha\ge 2}\sum_i (\delta P_{0i} {u}_{\alpha, i})^2 \frac{T}{\lambda_\alpha(\lambda_\alpha + \tau_0^{-1})} + \mathcal{O}(T^0) \, ,\\
\overline{{\mathcal{C}_2}}(T)  &=&\sum_{\alpha\ge 2}\sum_i (\delta P_{0i} {u}_{\alpha, i})^2 \frac{\tau_0^{-1} T}{(\lambda_\alpha + \tau_0^{-1})} + \mathcal{O}(T^0) \, .
\end{eqnarray}
Averaging over all permutations, $\sigma$, of the components of $\delta {\bm P}_0=(\delta P_{01}, ... , \delta P_{0n})$ , one has the following identity,
\begin{eqnarray}
\frac{1}{n!}\sum_\sigma\sum_i (\delta P_{0\sigma(i)} {u}_{\alpha, i})^2&=&\frac{(n-1)!}{n!}\left(\sum_i \delta P_{0i}^2\right)\left(\sum_j u_{\alpha,j}^2\right)=\frac{(\delta {P}_0)^2}{n}\equiv\langle \delta {{P}_0}^2 \rangle.
\end{eqnarray}
We finally obtain the leading-order contribution in $T$,
\begin{eqnarray}
\langle{{\mathcal{C}_1}}\rangle(T) &=&\langle\delta {P}_0^2\rangle\sum_{\alpha\ge 2} \frac{T}{\lambda_\alpha(\lambda_\alpha + \tau_0^{-1})} + \mathcal{O}(T^0) \, ,\\
\langle{{\mathcal{C}_2}}\rangle(T) &=&\langle\delta {P}_0^2\rangle\sum_{\alpha\ge 2} \frac{\tau_0^{-1} T}{(\lambda_\alpha + \tau_0^{-1})} + \mathcal{O}(T^0) \, ,
\end{eqnarray}
which can be Taylor-expanded in geometric series when either $\lambda_\alpha \tau_0 > 1$ or
$\lambda_\alpha \tau_0 < 1$, $\forall \alpha$ to obtain $\langle {\mathcal C}_{1,2}\rangle
= n^{-1} \, \langle \delta P_0^2 \rangle \, T \, \sum_{m=0}^{\infty} C_{1,2}^{(m)}$. One easily
recovers the coefficients $C_{1,2}^{(m)}$ given in Eqs.~(15a,b) in the main text. Note that, when $\lambda_\alpha\tau_0\ll 1$,
\begin{eqnarray}
\langle{{\mathcal{C}_1}}\rangle(T) &\simeq &\frac{T\langle\delta {P}_0^2\rangle}{ n} C_1^{(0)}=\frac{T\langle\delta {P}_0^2\rangle\tau_0}{ n}\Kf_1  \,, \\
\langle{\mathcal{C}_2}\rangle(T) &\simeq &\frac{T\langle\delta {P}_0^2\rangle}{n}C_2^{(0)} = \frac{T\langle\delta {P}_0^2\rangle}{n}\Kf_0 \, .
\end{eqnarray}
For $\lambda_\alpha\tau_0\gg 1$, one obtains,
\begin{eqnarray}
\langle{{\mathcal{C}_1}}\rangle(T) &\simeq &\frac{T\langle\delta {P}_0^2\rangle}{ n} C_1^{(0)}=\frac{T\langle\delta {P}_0^2\rangle}{n}\Kf_2  \,, \\
\langle{{\mathcal{C}_2}}\rangle(T) &\simeq &\frac{T\langle\delta {P}_0^2\rangle}{n}C_2^{(0)} = \frac{T\langle\delta {P}_0^2\rangle}{\tau_0 n}\Kf_1 \, .
\end{eqnarray}
\subsection{Perturbations as Fourier series}
From Eq.~(6) in the main text, it is clear that, with $c_\alpha(0)=0$, $\mathcal{C}_{1,2}$ will always be a sum over 
eigenmodes of $\mathbb L$ labeled $\alpha$, and that each term in that sum contains a factor
$(\delta {\bm P} \cdot {\mathbf u}_\alpha)^2$, regardless of the choice of perturbation. Our first main conclusion, 
that the response of the synchronous state under specific, nonaveraged perturbation depends on the spectrum of 
$\mathbb L$ and on the overlap of its eigenmodes with the perturbation vector is therefore rather general. 

Our approach can moreover be extended to any perturbation that can be expanded in a Fourier series, 
\begin{equation}
\delta {\bm P}(t) = \sum_p \Big(\delta {\bm P}_{p}^+ \exp[2 \pi i p t/\tau_0] + \delta {\bm P}_{p}^- \exp[-2 \pi i p t/\tau_0] \Big) \, .
\end{equation}
The condition that the components of the perturbation vector are real, $\delta {P}_i \in \mathbb R$, gives either $\delta P_{p,i}^+=\delta P_{p,i}^- \in \mathbb R$ or $\delta P_{p,i}^+=-\delta P_{p,i}^- \in i\mathbb R$. Because $\delta {\bm P}(t=0)=0$, 
we consider only the latter case in what follows.
Eq.~(6) in the main text gives
\begin{equation}
c_\alpha(t) = \exp[-\lambda_\alpha t] \sum_{p,i} u_{\alpha,i} \, \delta P_{p,i}(t) \Big(\frac{e^{(\lambda_\alpha+2 \pi i p/\tau_0) t}-1}{\lambda_\alpha+2 \pi i p/\tau_0} - \frac{e^{(\lambda_\alpha-2 \pi i p/\tau_0) t}-1}{\lambda_\alpha-2 \pi i p/\tau_0} \Big) \, .
\end{equation}
In the long time limit we obtain
\begin{equation}
c_\alpha(t \rightarrow \infty ) = \sum_{p,i} u_{\alpha,i} \, |\delta P_{p,i}(t)| \, \frac{(4 \pi p/\tau_0) \, \cos(2 \pi p t /\tau_0)-2 \lambda_\alpha \sin(2 \pi p t/\tau_0)}{\lambda_\alpha^2+4 \pi^2 p^2/\tau_0^2}
\end{equation}
To get the average of the performance measure $\mathcal C_1$, we
square this expressions, average it over an homogeneous ensemble of perturbation as in the main text and sum over $\alpha$.
For a sufficiently long duration of perturbation, integrating over
time gives the dominant contribution to the fragility performance measures (under the assumption that $T$ is large, but 
shorter than the duration of  the perturbation) 
\begin{equation}\label{eq:c1fourier}
{\mathcal C}_1(T) \simeq  \frac{T \, \tau_0^2}{2} \, \sum_{\alpha,p}  \langle \delta {\bm P}_{p,0}^2 \rangle \, \frac{4 \lambda_\alpha^2 \tau_0^2 + 16 \pi^2 p^2 
}{(\lambda_\alpha^2 \tau_0^2+4 \pi^2 p^2)^2} \, .
\end{equation}
For each Fourier harmonics, the denominator can be Taylor-expanded, depending on whether $\lambda_\alpha \tau_0>2 \pi p$
of $\lambda_\alpha \tau_0 <2 \pi p$. When $2 \pi p/\tau_0$ lies outside the spectrum of the Laplacian, 
Eq.~(\ref{eq:c1fourier}) allows to express ${\mathcal C}_1(T)$ as a sum over even-order generalized
Kirchhoff indices.

\section{Kirchhoff indices and phase dynamics}
\subsection{Kirchoff indices in the cycle model with nearest and $q^\textrm{th}$-neighbor coupling }
The eigenvalues $\lambda_\alpha$ of the Laplacian of our model with uniform nearest and $q^{\text{th}}$-neighbor 
coupling are obtained by a Fourier transformation and are  given by,
\begin{eqnarray}
\lambda_{\alpha}=4-2\cos(k_{\alpha})-2\cos(k_{\alpha}q) \; \;, \; \alpha=1,... ,n\; ,
\end{eqnarray} 
where $k_{\alpha}=\frac{2\pi (\alpha-1)}{n}$. Then one obtains,
\begin{eqnarray}\label{eq:kf1}
\Kf_1=n\sum_{\alpha\ge 2}\frac{1}{4-2\cos(k_{\alpha})-2\cos(k_{\alpha}q)} \; \; .
\end{eqnarray}
The dependence of the denominator of Eq.~(\ref{eq:kf1}) with $q$ makes it is clear that $\Kf_1$ is a non monotonous
function of $q$. This is shown in Fig.~\ref{fig:fig2} for $n=50$. As mentioned in the main text, two seemingly similar choices of long range interactions may lead 
to large differences of the Kirchhoff index.
This translates into large variations of the network's average resistance distance, which can be understood topologically in terms of the commensurability of the $q^\textrm{th}$-neighbor coupling with 
the number $n$ of nodes in the network.
Since the resistance distance between any two nodes accounts for all paths between them, one may expect that
$q^\textrm{th}$-neighbor couplings provide short, alternative paths between nodes, effectively bringing them closer to each other.
This is however not always the case. In fact, if $n$ is a small integer multiple of $q$, or nearly so, 
paths involving multiple $q$-range hops, starting from a given node, come back to the initial node or close to it.
Such paths only allow to reach nodes in the close vicinity of the starting node and do not reduce significantly the 
resistance distance between many nodes close to the original one. 
In contrast, without commensurability between $q$ and $n$, the resistance distance between most of the nodes to the original one is reduced.

This is illustrated in Fig.~\ref{fig:fig2} which plots $\Kf_1$ and $\Kf_2$ as a function of $q$ for $n=50$.
As expected peaks are present for $q = 10, 17$ and $24$ such that $n/q$ is a small integer or close to a small integer. The insets of Fig.~\ref{fig:fig2} sketch 
how despite these long range interactions some portions of the network keep the same geodesic distance to the red node $1$.

\begin{figure}
\centering
\includegraphics[width=1\columnwidth]{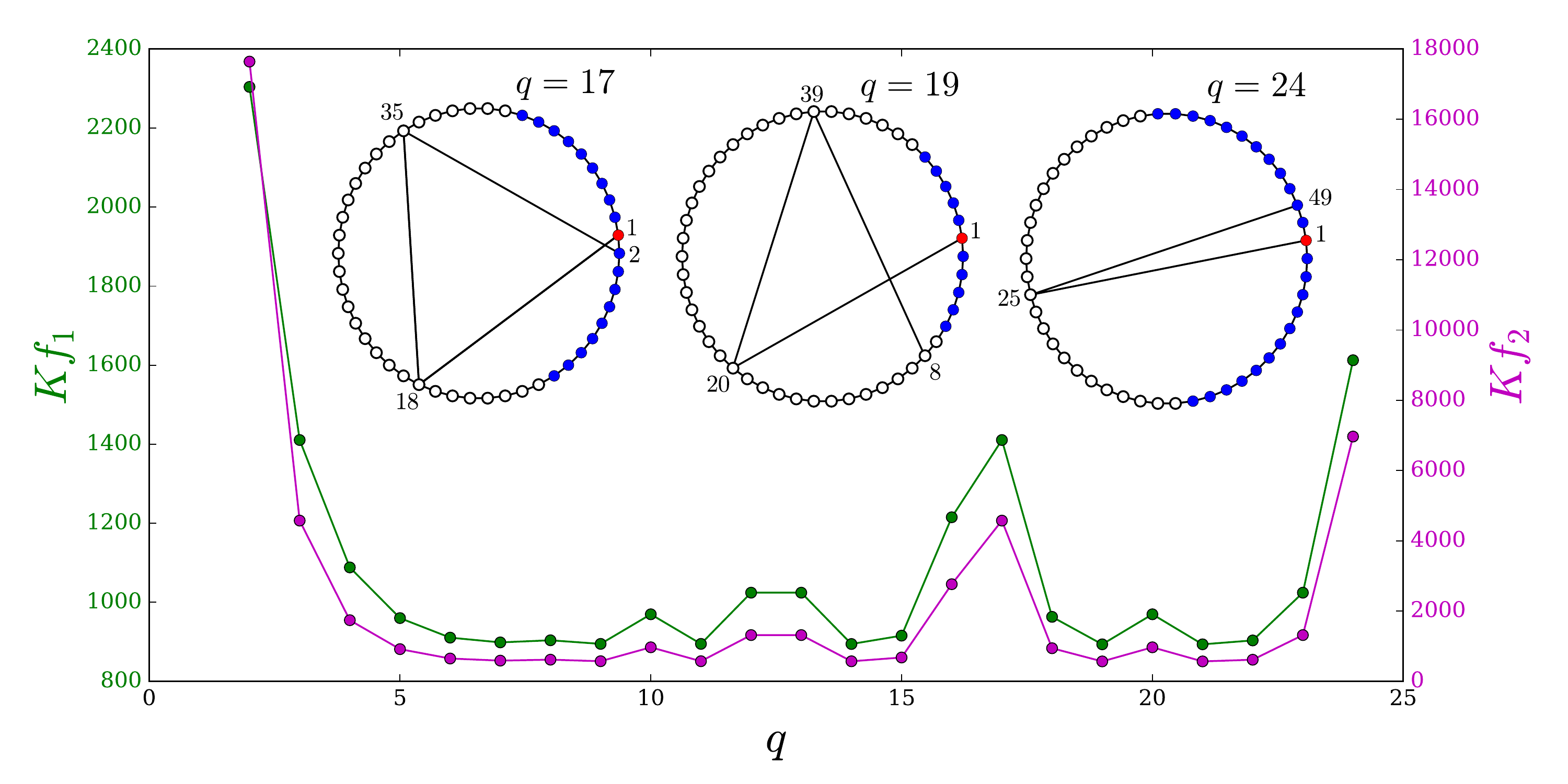}
\caption{\label{fig:fig2}(Color online) $\Kf_1$ (green) and $\Kf_2$ (violet) for a cyclic graph with $n=50$ 
with nearest and $q^{\text{th}}$-neighbor coupling, $b_{i,i\pm1}=b_{i,i\pm q}=b_0$. 
The inset sketches the model for $q=17$, $19$, and $24$ and illustrates one path involving $q^\textrm{th}$-range 
interactions starting from node $1$ (red). 
The addition of $q^\textrm{th}$- neighbor interactions does not reduce the geodesic distance between the reference node
(red) and the set of nodes colored in blue.}
\end{figure}

\subsection{Influence on the phase dynamics} 
The two seemingly similar cycle models with $n=50$, nearest- and $q^{\rm th}$-neighbor coupling with 
$q=17$ and $q=19$ considered in Fig.~2 in the main text nevertheless behave
strikingly differently under an external perturbation. This is so, because they have very different generalized Kirchhoff indices $\Kf_1$ and $\Kf_2$.
To illustrate this behavioral discrepancy, 
Fig.~\ref{fig:fig1} compares the phase dynamics for the box perturbation for these two graphs. The left panel is for the cyclic graph with
$q=17$, which has a fragility performance measure $\mathcal{C}_1$ bigger than for the cyclic graph with $q=19$ in the right panel. Clearly the angle deviations on the left panel spread more and take more time to return to the initial fixed point after the 
perturbation than on the right panel. From Eq.~(13b) in the main text, 
$\langle {\mathcal C}_1^\infty\rangle $ is proportional to $\Kf_2$ in the corresponding limit $\lambda_\alpha \tau_0 \gg 1$ $\forall \alpha$. 
The numerically obtained values (indicated in Fig.~\ref{fig:fig1}) of $\mathcal{C}_1$ 
follow that trend, though not exactly, as expected for this single realization of perturbation.
\begin{figure}
\centering
\includegraphics[width=1\columnwidth]{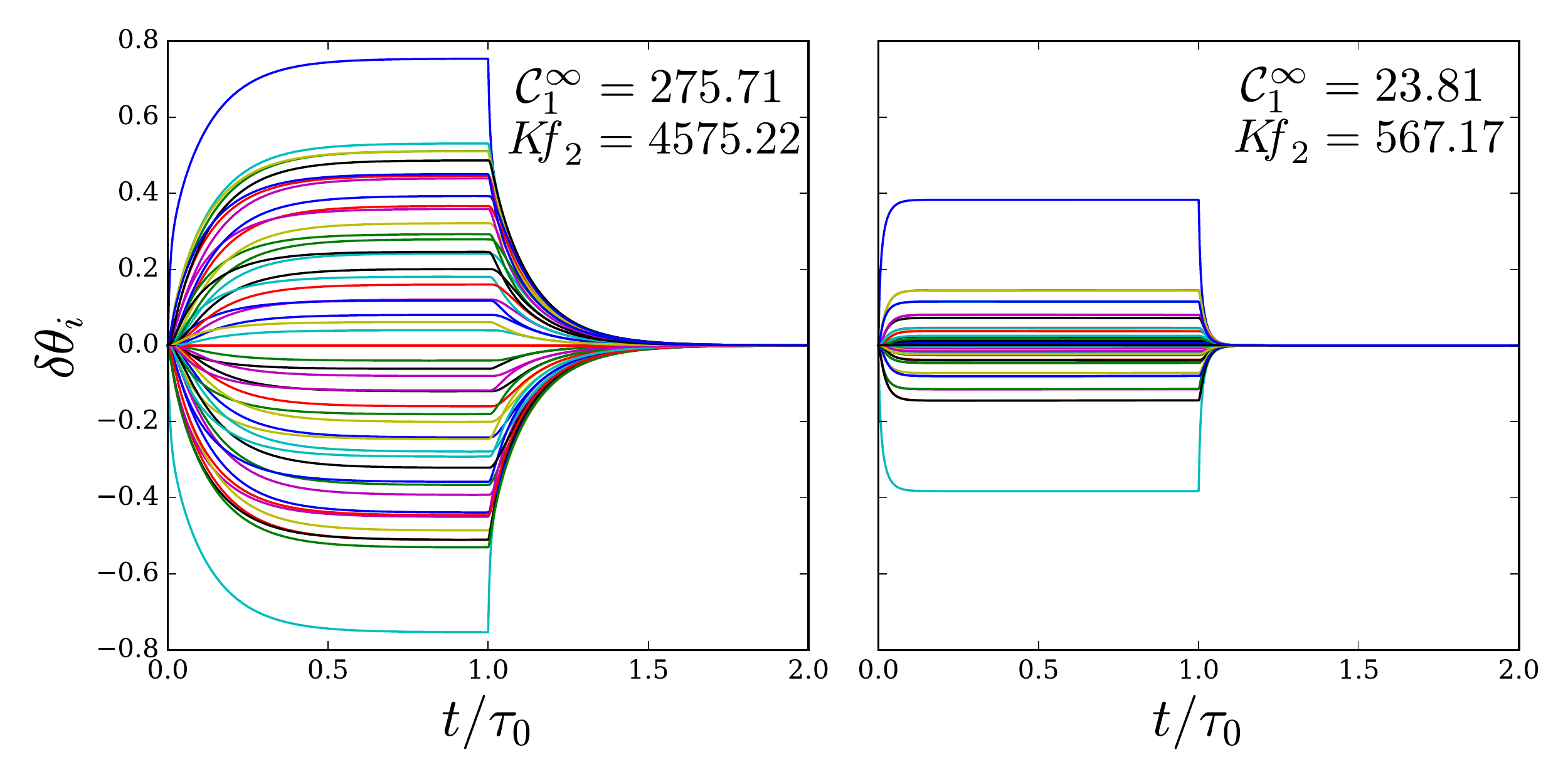}
\caption{\label{fig:fig1} (Color online) Phases $\delta \theta_i$ for the cyclic graph with $n=50$ with nearest and $q^{\text{th}}$-neighbor coupling, $b_{i,i\pm1}=b_{i,i\pm q}=b_0$ with $q=17$ (left panel) and $q=19$ (right panel), as a function of the normalized time $t/\tau_0$, for a box perturbation with $\tau_0=50/b_0$ and perturbation vector with non zero components $\delta P_{0,1}=b_0$, $\delta P_{0,11}=-b_0$.}
\end{figure}

\subsection{$\lambda_2$ vs. Generalized Kirchhoff indices}
\begin{figure}
\centering
\hspace{-3mm}\includegraphics[width=1.02 \columnwidth]{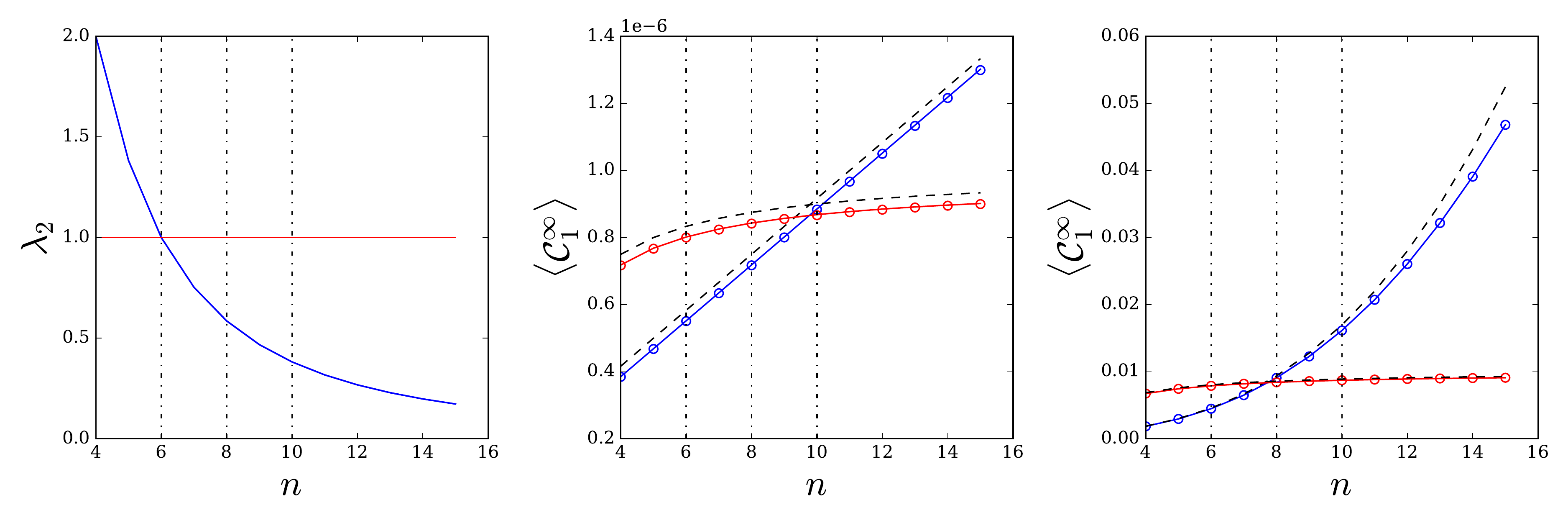}
\caption{\label{fig:fig3} (Color online) $\lambda_2$ (left panel) and performance measure $\langle {\mathcal{C}}_1^\infty \rangle$ for a box perturbation with $\tau_0=50/b_0$ (center panel) and $\tau_0=0.1/b_0$ (right panel) as a function of the number of nodes for the star network (red) and the cyclic network (blue). Solid lines give Eq.~(13a) of the main text. The dashed lines give the two limits of Eq.~(13a) : $\lambda_\alpha\tau_0\ll 1$ (center panel) and $\lambda_\alpha\tau_0\gg 1$ (right panel)}
\end{figure}

Here we show that the generalized Kirchoff indices give more information on the fragility of synchronous states
than the smallest Lyapunov exponent, $\lambda_2$. 
We compare star and cycle graphs with the same number of nodes. In both cases
the eigenvalues of the  Laplacian matrix can be calculated analytically. 
The spectrum of the Laplacian of a star graph with $n$ nodes is $\{0,1,n\}$, with the eigenvalue $1$
having multiplicity $n-2$, thus $\lambda_2=1$, $\Kf_1=(n-1)^2$,
and $\Kf_2=(n^3-2n^2-1)/n$. 
The spectrum of the Laplacian of a cycle graph with only nearest neighbor coupling is 
$\lambda_\alpha=2[1-\cos(2\pi(\alpha-1)/n)]$, with $\alpha = 1, \ldots, n$.
One has, $\lambda_2=2[1-\cos(2\pi/n)]$, $\Kf_1=n(n^2-1)/12$, and $\Kf_2=n(n^2-1)(n^2+11)/720$. 
The density of eigenvalues approaching zero increases with $n$ in the cycle graph, while
in the star graph the Lyapunov exponents accumulate at a finite value as $n$ increases 
(i.e. $\lambda_2=\lambda_3=\ldots=\lambda_{n-1}=1$). 
Thus we expect a crossover in the vulnerability of these two network topologies as $n$ is increased.
Fig.~\ref{fig:fig3} shows $\lambda_2$ (left panel), 
and the performance measure ${\mathcal{C}}_1$ 
in both the limits $\lambda_\alpha\tau_0\ll 1$ (center panel) and $\lambda_\alpha\tau_0\gg 1$ (right panel) 
as a function of the number of nodes. 
For $n<6$, the cycle network has a larger $\lambda_2$ and a smaller ${\mathcal{C}}_1$ compared to the star 
network which means that the largest contribution to $\Kf_1$ and $\Kf_2$ comes from $\lambda_2^{-1}$. 
However, for $6>n>8$, both $\Kf_1$ and $\Kf_2$ are not dominated by $\lambda_2^{-1}$, therefore the cycle network is less 
fragile under an external perturbation than the star network even though it has a smaller $\lambda_2$. 
For $n=8$ and $n=9$, the cycle network is less fragile against short time perturbation but more fragile against long 
time perturbation compared to the star network. This reflects the fact that for those values, $\Kf_1$ is smaller for the cycle
than for the star network, while the relation is opposite for $\Kf_2$ [see the corresponding  
relation between $\mathcal C_1$ and generalized Kirchhoff indices in Eq.~(13a) in the main text].

\begin{figure}
\centering
\hspace{-3mm}\includegraphics[width=1.02 \columnwidth]{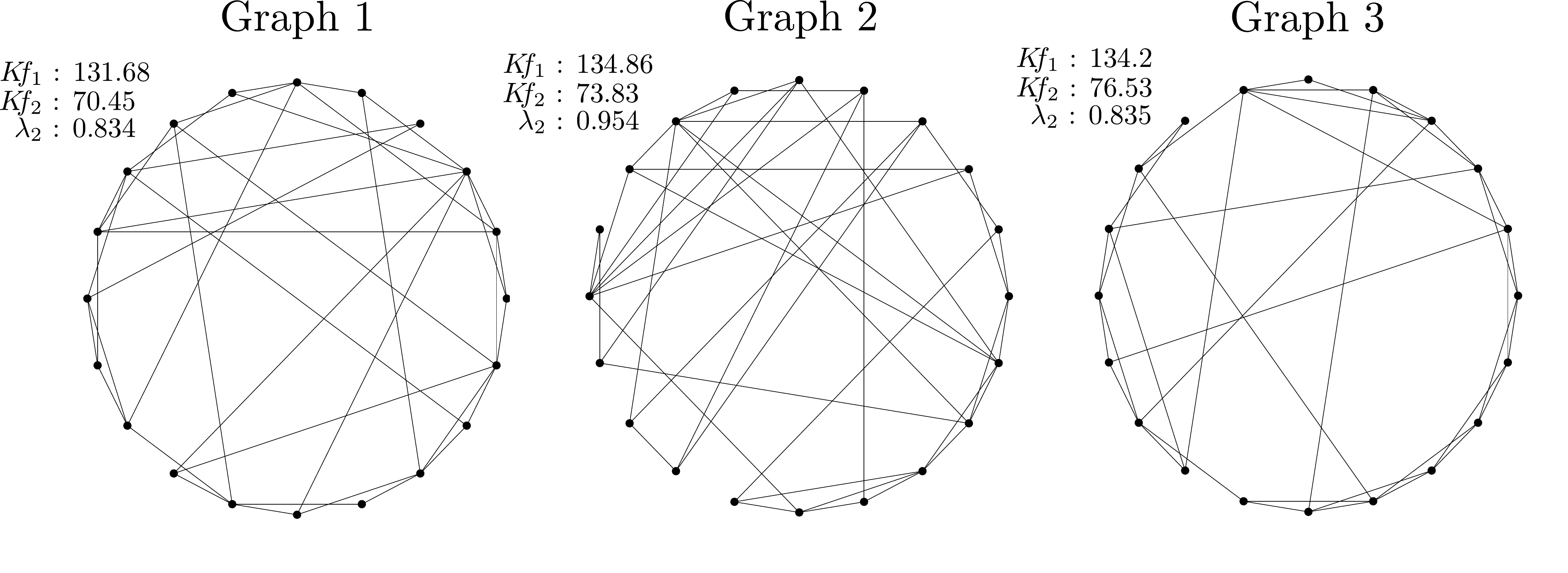}
\caption{\label{fig:fig4} Starting from a homogeneous cyclic network with first and second nearest 
neighbor couplings and size $n=20$, graphs 1, 2 and  3 have been obtained after rewiring some randomly chosen edges.}
\end{figure}

We apply the same analysis to small-world graphs, which are obtained from 
a $n=20$ cycle network with first and second nearest neighbor
couplings, which are rewired~\cite{Wat98}.
Fig.~\ref{fig:fig4} shows three graphs obtained with this procedure, which have different relations between $\lambda_2$
and their generalized Kirchhoff indices $\Kf_1$ and $\Kf_2$.
Graph 1 has a smaller $\lambda_2$ but smaller $\Kf_1$ or $\Kf_2$ compared to graph 2, while
graph 3 has a smaller $\lambda_2$, similar $\Kf_1$ and larger $\Kf_2$ compared to graph 2. 
These relations are reflected by the performance measures ${\mathcal{C}}_1$ for box perturbations 
presented in Table~\ref{tab:tab1}. 

\begin{center}
\begin{table}
\begin{tabular}{|c|c|c|c|}
  \hline
    \hline
   Graph & 1 & 2 & 3\\
  \hline
  \hline
  $\lambda_2$ & 0.834 & 0.954 & 0.835\\
$\lambda_\alpha \tau_0 \ll 1$ : $\langle{\mathcal{C}}_1^\infty \rangle\propto \Kf_1$ & 3.16e-7 & 3.24e-7 & 3.22e-7 \\
  $\lambda_\alpha \tau_0 \gg 1$ : $\langle{\mathcal{C}}_1^\infty \rangle\propto \Kf_2$ & 1.83e-3 & 1.91e-3 & 1.98e-3 \\
  \hline
\end{tabular}
\caption{$\lambda_2$, performance mesure $\langle{\mathcal{C}}_1^\infty\rangle$ obtained numerically in the two limits : $\lambda_\alpha \tau_0 \ll 1$ ($\tau_0=0.1/b_0$), $\lambda_\alpha \tau_0 \gg 1$ ($\tau_0=50/b_0$).}\label{tab:tab1}
\end{table}
\end{center}

\section{Beyond Kuramoto}
Instead of the Kuramoto model defined by Eq.~(1) in the main text, one may consider other models of coupled dynamical systems.
Two extensions have to be differentiated. First, one may consider different coupling than the sine-coupling in the Kuramoto model. 
This is straightforwardly included in our approach and leads only to a differently weighted Laplacian matrix, giving, for a fixed network,
a different Lyapunov spectrum and different eigenmodes ${\mathbf u}_\alpha$, but leaving all expressions for the performance measures
unchanged. Second, one may consider dynamical systems with more internal degrees of freedom, such as the one considered
by Pecora and Carroll~\cite{Pec98}
\begin{equation}
\dot{\mathbf x} = {\mathbf P}({\mathbf x}) + {\mathbb B} \otimes {\mathbf H}({\mathbf x}) \, ,
\end{equation}
with ${\mathbf x} = ({\mathbf x}_1, {\mathbf x}_2, ... {\mathbf x}_n)$, ${\mathbf P}({\mathbf x})=\big ({\mathbf f}({\mathbf x}_1),
{\mathbf f}({\mathbf x}_2), ... {\mathbf f}({\mathbf x}_n) \big)$, $\mathbb B$ the Laplacian matrix of the graph considered 
and $\mathbf H$ a function defining the coupling between adjacent dynamical systems  with 
coordinates ${\mathbf x}_i \in {\mathbb R}^d$. Our approach assumes the existence of 
a synchronous state with ${\mathbf x}_1^{(0)}={\mathbf x}_2^{(0)}=... ={\mathbf x}_n^{(0)}$. Linearizing about the synchronous state with 
${\mathbf x}={\mathbf x}^{(0)}+\delta {\mathbf x}$ and considering a perturbation gives, instead of Eq.~(2) in the main text, 
\begin{equation}
\delta \dot {\mathbf x} =\delta {\mathbf P} + [{\mathbb I} \otimes D{\mathbf P}({\mathbf x}^{(0)})+ {\mathbb B} \otimes D {\mathbf H}({\mathbf x}^{(0)}) ] \, \delta {\mathbf x} \, ,
\end{equation}
where $D{\mathbf P}$ and $D {\mathbf H}$ are Jacobian matrices.
The first term on the right-hand side is similar to the perturbation considered in the main text, and the third one
is a generalization of the Laplacian term in Eq.~(2) in the main text, where the network Laplacian is extended to take account of 
additional nodal degrees of freedom. The new second term occurs because ${\mathbf P}$ now
depends on internal degrees of freedom ${\mathbf x}$ (it does not in the Kuramoto model). The formula given in the main text
for the performance need now to be evaluated with the eigenvalues $\Lambda_{\alpha,l}$ and eigenmodes 
${\mathbf U}_{\alpha,l}$ of 
${\mathbb I} \otimes D{\mathbf P}({\mathbf x}^{(0)})+{\mathbb B} \otimes D {\mathbf H}({\mathbf x}^{(0)})$, $l=1,2,... d$.

\end{document}